\documentclass{iau} 
\usepackage{graphicx,subfigure}

\title[ULXs as super-critical propellers and accretors in HMXBs] %% give here short title %%
{Ultra-luminous X-ray sources \\ as neutron stars propelling and accreting \\ at super-critical rates \\ in high-mass X-ray binaries}

\author[M. H. Erkut \& K. Y. Ek{\c s}i]   %% give here short author list %%
{M. Hakan Erkut$^{1,a}$ \and K. Yavuz Ek{\c s}i$^{1,b}$}

\affiliation{$^1$Physics Engineering Department, Faculty of Science and Letters, \\ Istanbul Technical University, 34469, Istanbul, Turkey \\ $^a$email: {\tt mherkut@gmail.com}, $^b$email: {\tt eksi@itu.edu.tr}}

\pubyear{2018}
\volume{346}  %% insert here IAU Symposium No.
\jname{High-mass X-ray binaries: illuminating the passage from massive binaries to merging compact objects}
\editors{A.C. Editor, B.D. Editor \& C.E. Editor, eds.}
\begin{document}

\maketitle

\begin{abstract}
Ultra-luminous X-ray sources (ULXs) are off-nuclear point sources in nearby galaxies with luminosities well exceeding the Eddington limit for stellar-mass objects. It has been recognized after the discovery of pulsating ULXs (PULXs) that a fraction of these sources could be accreting neutron stars in high-mass X-ray binaries (HMXBs) though the majority of ULXs are lacking in coherent pulsations. The earliest stage of some HMXBs may harbor rapidly rotating neutron stars propelling out the matter transferred by the massive companion. The spin-down power transferred by the neutron-star magnetosphere to the accretion disk at this stage can well exceed the Eddington luminosities and the system appears as a non-pulsating ULX. In this picture, PULXs appear as super-critical mass-accreting descendants of non-pulsating ULXs. We present this evolutionary scenario within a self-consistent model of magnetosphere-disk interaction and discuss the implications of our results on the spin and magnetic field of the neutron star.
\end{abstract}

\firstsection % if your document starts with a section,
\section{Introduction}

A significant fraction of ultra-luminous X-ray sources (ULXs) may consist of neutron stars as indicated by the recent detection of pulsations from M82~X-2 (\cite[Bachetti et al. 2014]{Bachetti_etal2014}), ULX~NGC~7793~P13 (\cite[F{\"u}rst et al. 2016]{Furst_etal2016}), ULX~NGC~5907 (\cite[Israel et al. 2017]{Israel_etal2017}), and NGC~300~ULX1 (\cite[Carpano et al. 2018]{Carpano_etal2018}). In addition to pulsating ULXs (PULXs), ultra-luminous super-soft sources (ULSs) (\cite[Di Stefano \& Kong 2003]{DiStefanoKong2003}; \cite[Fabbiano et al. 2003]{Fabbiano_etal2003}; \cite[Kong \& Di Stefano 2003]{KongDiStefano2003}) and other non-pulsating ULXs emerge as seemingly different subclasses of the ULX population. The lack of pulsations can be due to the optically thick envelope fed by the outflows of the accretion disk around the neutron star (\cite[Ek{\c s}i et al. 2015]{Eksi_etal2015}) and/or the propeller effect (\cite[Illarionov \& Sunyaev 1975]{IllarionovSunyaev1975}) of the neutron-star magnetosphere on the disk matter (\cite[Tsygankov et al. 2016]{Tsygankov_etal2016}).

We consider the spin and luminosity evolution of neutron stars in high-mass X-ray binaries (HMXBs). The magnetosphere of the newborn rapidly rotating neutron star with spin periods of a few milliseconds interacts with the wind-fed disk in the very early stage of the X-ray binary (\cite[Erkut et al. 2018]{Erkut_etal2018}). The donor, as an already evolved massive star, produces dense winds with high mass-loss rates $(\dot{M}_{\rm w}\gtrsim 10^{-6}\, M_\odot \, {\rm yr}^{-1})$. Such an evolutionary scheme can be illustrated by a neutron star--helium star binary that is expected to form soon after the common-envelope phase of a twin massive binary (\cite[Brown 1995]{Brown1995}; \cite[Dewi et al. 2006]{Dewi_etal2006}). It takes $\sim 10^6$~years for the helium-burning stage to end. During its lifetime ($\sim 10^6$~years), the massive helium star loses mass at a rate of $\dot{M}_{\rm w}\sim 10^{-5}\, M_\odot \, {\rm yr}^{-1}$. Following the fusion of elements heavier than helium within $\lesssim 10^4$~years, the helium-star core collapses to give rise to the birth of the second neutron star. Double neutron-star systems might therefore be the descendants of helium star--neutron star X-ray binaries (\cite[Dewi et al. 2006]{Dewi_etal2006}).

\begin{figure}[htb]
% \vspace*{-2.0 cm}
\begin{center}
 \includegraphics[width=4.3in]{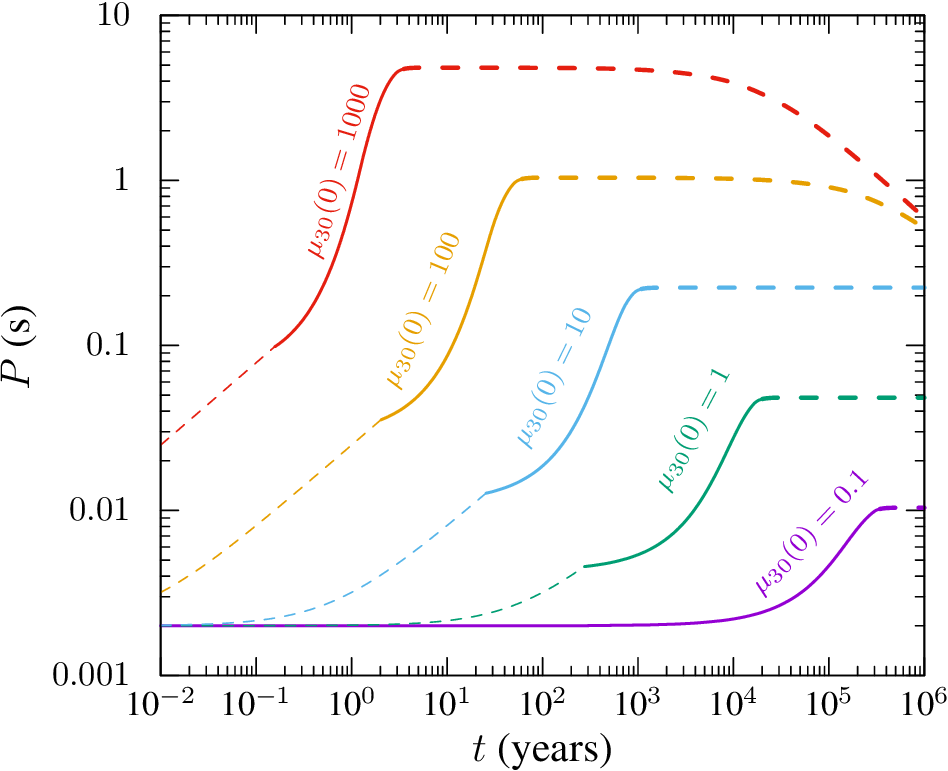} 
% \vspace*{-1.0 cm}
 \caption{Spin-period evolution with super-critical ejector (thin dashed), propeller (solid), and accretor (thick dashed) phases for a set of initial magnetic moments. We assume that the dipole magnetic fields stronger than $B=10^9\, {\rm T}$~($\mu_{30}=10$ at $t=0$) decay according to the scenario B (\cite[Erkut et al. 2018]{Erkut_etal2018}).}
   \label{fig1}
\end{center}
\end{figure}

\section{Model}

The fraction of the mass-loss rate of a $\sim 20\, M_\odot$ wind donor to be captured by a $1.4\, M_\odot$ neutron star can be as high as $\dot{M}_0/\dot{M}_{\rm w}\sim 0.3$ due to photoionization of the wind matter irradiated by the X-rays emitted from the neutron star. Deceleration of the wind matter usually leads to the formation of an extensive disk around the neutron star with high mass-transfer rates (\cite[\v{C}echura \& Hadrava 2015]{CechuraHadrava2015}).

In our picture, the mass transfer from the massive helium companion (Wolf-Rayet star) with mass-loss rates of $\dot{M}_{\rm w}\gtrsim 10^{-5}\, M_\odot \, {\rm yr}^{-1}$  to the neutron star occurs at super-Eddington (super-critical) rates $(\dot{M}_0\gtrsim 10^{-6}\, M_\odot \, {\rm yr}^{-1})$. The innermost disk radius (magnetopause) is smaller than the spherization radius, i.e., $R_{\rm in} < R_{\rm sp}$. The system is thus in the super-critical regime (\cite[Shakura \& Sunyaev 1973]{ShakuraSunyaev1973}). The neutron star acts as a super-critical propeller for $R_{\rm co} < R_{\rm in} < R_{\rm sp}$. The super-critical accretion regime is realized when $R_{\rm in} < R_{\rm co} < R_{\rm sp}$. The corotation radius in the disk, $R_{\rm co}\equiv (GM/\Omega_*^2)^{1/3}$, is determined by the neutron-star spin period, $P=2\pi/\Omega_*$. The ejector phase is realized when $R_{\rm L} < R_{\rm in}$, where $R_{\rm L}=c/\Omega_*$ is the light-cylinder radius.

\begin{figure}[htb]
\centering
\subfigure{\includegraphics[width=4.5in]{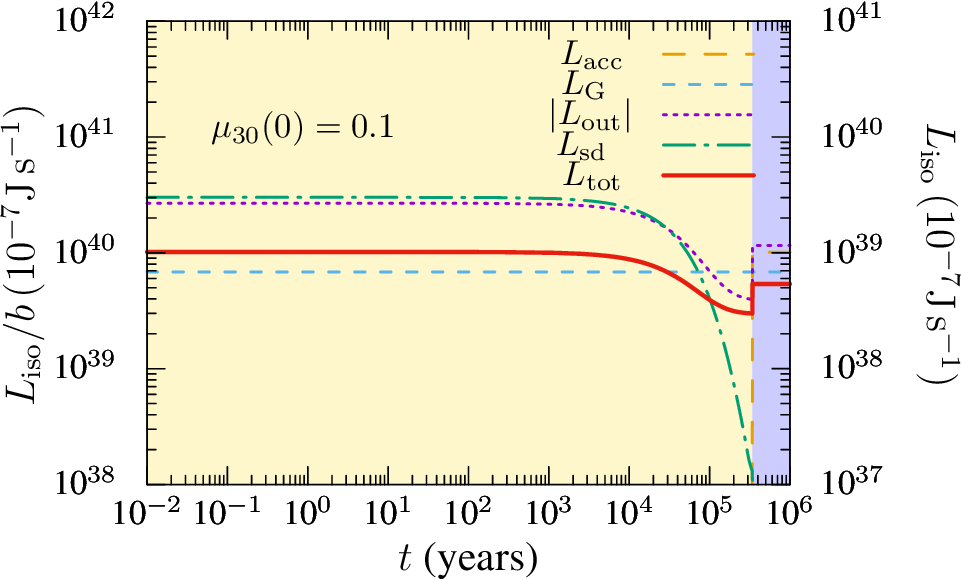}}
\subfigure{\includegraphics[width=4.5in]{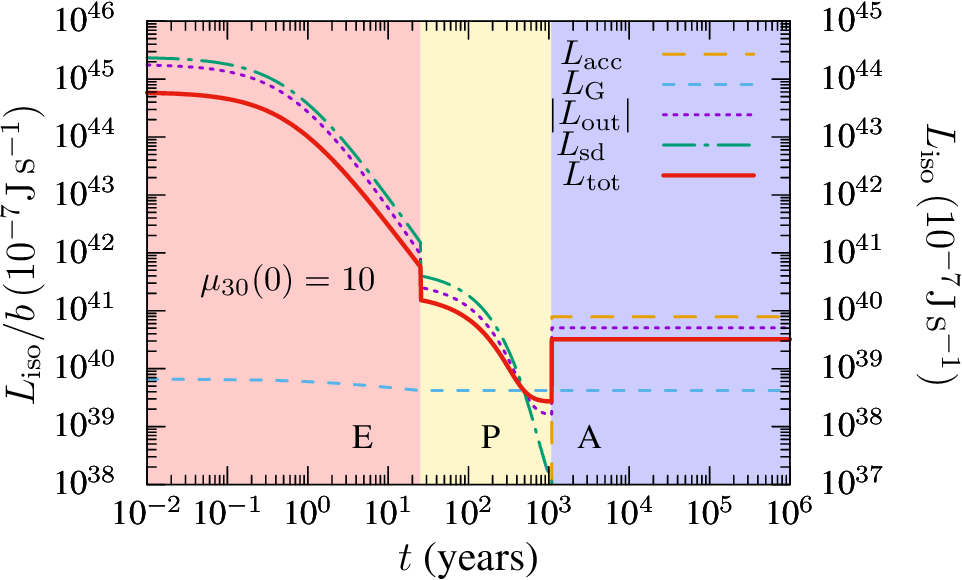}}
\caption{Luminosity evolution for two different initial magnetic moments. The luminosity for isotropic emission is on the right vertical axis. The left vertical axis represents the observed luminosity if the geometrical beaming is $b=0.1$. Shaded regions correspond to phases such as ejector (E), propeller (P), and accretor (A) (\cite[Erkut et al. 2018]{Erkut_etal2018}).}
\label{fig2}
\end{figure}

To obtain the spin-period evolution of the neutron star of moment of inertia $I$, we solve the torque equation,
\begin{equation}
-\frac{2\pi I}{P^2}\frac{dP}{dt}=N. \label{trqeqn}
\end{equation}
In the ejector phase, the torque, $N$, acting on the neutron star is due to the magnetic dipole radiation, i.e., $N\simeq-(2/3)\mu^2\Omega_*^3/c^3$. For the super-critical propeller and accretion regimes, we write
\begin{equation}
N=n\left(\omega_*\right)\dot{M}_{\rm in}\sqrt{GMR_{\rm in}} \label{torque}
\end{equation}
with the fastness parameter $\omega_* \equiv P_{\rm K,in}/P$ in terms of the mass inflow rate and the Keplerian period at $R_{\rm in}$. As a function of the fastness parameter, the dimensionless torque becomes $n < 0$ and $n > 0$ for the propeller and accretion regimes, respectively. We use $n\simeq 1$ for the accretion torque and $n\simeq1-\omega_*$ for the propeller torque (\cite[Erkut et al. 2018]{Erkut_etal2018}).

The total luminosity of the disk around a propeller with subcritical mass-inflow rates can be written as $L_{\rm tot}=GM\dot{M}/R_{\rm in}-I\Omega_*\dot{\Omega}_*-\dot{M}_{\rm out} v_{\rm out}^2/2$ (\cite[Ek\c{s}i et al. 2005]{Eksi_et2005}). In the super-critical propeller regime, however, each term contributing to the total luminosity of the neutron star--disk system must be treated accordingly by taking into account the regulation of the accretion flow inside the spherization radius. Noting that the mass-inflow rate, throughout the disk, is $\dot{M}=\dot{M}_0(R/R_{\rm sp})$ for $R<R_{\rm sp}$ and $\dot{M}=\dot{M}_0$ for $R>R_{\rm sp}$ (\cite[Shakura \& Sunyaev 1973]{ShakuraSunyaev1973}), we calculate the total luminosity of the neutron-star-disk system in the super-critical accretion regime using $L_{\rm tot} = L_{\rm acc} + L_{\rm out} + L_{\rm G}$ and in the super-critical ejector and propeller regimes using $L_{\rm tot} = L_{\rm sd} + L_{\rm out} + L_{\rm G}$. As a result of super-critical mass inflow, $L_{\rm out } < 0$ represents the energy-loss rate due to outflows from the disk. The rate of gravitational energy release throughout the disk is given by $L_{\rm G}$. The spin-down power and accretion luminosity can be written as $L_{\rm sd}=-2\pi N/P$ and $L_{\rm acc}=GM\dot{M}_{\rm in}/R_*$ with $\dot{M}_{\rm in}=\dot{M}_0(R_{\rm in}/R_{\rm sp})$, respectively, for a neutron star of mass $M$ and radius $R_*$ (\cite[Erkut et al. 2018]{Erkut_etal2018}).

\section{Results}

In the super-critical propeller and accretion regimes, the inner disk radius depends on several parameters, that is,
\begin{equation}
R_{\rm in}\simeq \left(\frac{\mu^2 R_{\rm sp} \delta}{\dot{M}_0\sqrt{GM}}\right)^{2/9}, \label{rin}
\end{equation}
where $\mu$, $R_{\rm sp}$, $\delta$, and $\dot{M}_0$ are the neutron-star magnetic dipole moment, the spherization radius, the width of magnetosphere-disk interaction zone, and the mass inflow rate in the outer disk (\cite[Erkut et al. 2018]{Erkut_etal2018}). We assume $M=1.4M_\odot$, $R_*=10\, {\rm km}$, $\dot{M}_0=4\times 10^{17}\, {\rm kg\, s^{-1}}(\sim 6\times 10^{-6}\, M_\odot \, {\rm yr^{-1}})$, $\delta =0.01$, and $P_0=2\, {\rm ms}$ for the initial period of the neutron star. We allow the field decay for initial magnetic field strengths in the magnetar range. Here, we present our results (Fig.\, \ref{fig1} and Fig.\, \ref{fig2}) for the field-decay mechanism B (\cite[Colpi et al. 2000]{Colpi_etal2000}) as an illustrative example (the other field mechanisms such as A and C yield similar equilibrium periods with different timescales).

\section{Conclusions}

As shown in Figure~\ref{fig1}, the observed spin periods of PULXs $(P\sim 1\, {\rm s})$ can be realized for sufficiently strong initial magnetic fields in the $B\sim 10^{9}-10^{11} \, {\rm T}\, (10^{13}-10^{15} \, {\rm G})$ range. In the very early (ejector) stage of the luminosity evolution, neutron stars with such strong initial fields can even appear as supernova impostors (lower panel of Fig.\, \ref{fig2}). The super-critical propeller stage, during which the source luminosity becomes comparable with those of ULXs, is much shorter for strongly magnetized neutron stars than for weakly magnetized neutron stars (Fig.\, \ref{fig2}). Neutron stars with relatively strong magnetic fields spend most of their lives in the super-critical accretion regime. It is therefore more likely that the neutron stars of $B > 10^{9}\, {\rm T}$ appear as PULXs (Fig.~\ref{fig1} and lower panel of Fig.~\ref{fig2}).

Most of the non-pulsating ULXs/ULSs may consist of neutron stars in the super-critical propeller regime (upper panel of Fig.~\ref{fig2}) with relatively weak magnetic fields $(B\sim 10^{7}\, {\rm T})$ and shorter (but hardly observable) spin periods $(P\sim 0.01\, {\rm s})$. Although the equilibrium periods of the weak-field ULXs are smaller than the observed typical periods of PULXs, the population of the weak-field systems can be larger than the population of PULXs. Yet, it would relatively be more difficult, due to the smaller size of the magnetosphere, to observe pulsations from these weakly magnetized neutron-star ULXs/ULSs.


\begin{thebibliography}{}

\bibitem[Bachetti \etal\ (2014)]{Bachetti_etal2014}
{Bachetti, M., Harrison, F.A., Walton, D.J., Grefenstette, B.W., Chakrabarty, D., Fürst, F., Barret, D., Beloborodov, A., Boggs, S.E., Christensen, F.E., Craig, W.W., Fabian, A.C., Hailey, C.J., Hornschemeier, A., Kaspi, V., Kulkarni, S.R., Maccarone, T., Miller, J.M., Rana, V., Stern, D., Tendulkar, S.P., Tomsick, J., Webb, N.A., \& Zhang, W.W.} 2014,
\textit{Nature}, 514, 202

\bibitem[Brown (1995)]{Brown1995}
{Brown, G.E.} 1995,
\textit{ApJ}, 440, 270

\bibitem[Carpano \etal\ (2018)]{Carpano_etal2018}
{Carpano, S., Haberl, F., Maitra, C., \& Vasilopoulos, G.} 2018,
\textit{MNRAS} (Letters), 476, L45

\bibitem[\v{C}echura \& Hadrava (2015)]{CechuraHadrava2015}
{\v{C}echura, J., \& Hadrava, P.} 2015,
\textit{A\&A}, 575, A5

\bibitem[Colpi \etal\ (2000)]{Colpi_etal2000}
{Colpi, M., Geppert, U., \& Page, D.} 2000,
\textit{ApJ} (Letters), 529, L29

\bibitem[Dewi \etal\ (2006)]{Dewi_etal2006}
{Dewi, J.D.M., Podsiadlowski, P., \& Sena, A.} 2006,
\textit{MNRAS}, 368, 1742

\bibitem[Di Stefano \& Kong (2003)]{DiStefanoKong2003}
{Di Stefano, R., \& Kong, A.K.H.} 2003,
\textit{ApJ}, 592, 884

\bibitem[Ek{\c s}i \etal\ (2005)]{Eksi_etal2005}
{Ek{\c s}i, K.Y., Hernquist, L., \& Narayan, R.} 2005,
\textit{ApJ} (Letters), 623, L41

\bibitem[Ek{\c s}i \etal\ (2015)]{Eksi_etal2015}
{Ek{\c s}i, K.Y., Anda{\c c}, {\.I}.C., {\c C}{\i}k{\i}nto{\u g}lu, S., Gen{\c c}ali, A.A., G{\"u}ng{\"o}r, C., \& {\"O}ztekin, F.} 2015,
\textit{MNRAS} (Letters), 448, L40

\bibitem[Erkut \etal\ (2018)]{Erkut_etal2018}
{Erkut, M.H., Ek{\c s}i, K.Y., \& Alpar, M.A.} 2018,
\textit{ApJ} (Submitted) eprint arXiv:1708.04502

\bibitem[Fabbiano \etal\ (2003)]{Fabbiano_etal2003}
{Fabbiano, G., King, A.R., Zezas, A., Ponman, T.J., Rots, A., \& Schweizer, F.} 2003,
\textit{ApJ}, 591, 843

\bibitem[F{\"u}rst \etal\ (2016)]{Furst_etal2016}
{F{\"u}rst, F., Walton, D.J., Harrison, F.A., Stern, D., Barret, D., Brightman, M., Fabian, A.C., Grefenstette, B., Madsen, K.K., Middleton, M.J., Miller, J.M., Pottschmidt, K., Ptak, A., Rana, V., \& Webb, N.} 2016,
\textit{ApJ} (Letters), 831, L14

\bibitem[Illarionov \& Sunyaev (1975)]{IllarionovSunyaev1975}
{Illarionov, A.F., \& Sunyaev, R.A.} 1975,
\textit{A\&A}, 39, 185

\bibitem[Israel \etal\ (2017)]{Israel_etal2017}
{Israel, G.L., Belfiore, A., Stella, L., Esposito, P., Casella, P., De Luca, A., Marelli, M., Papitto, A., Perri, M., Puccetti, S., Castillo, G.A.R., Salvetti, D., Tiengo, A., Zampieri, L., D'Agostino, D., Greiner, J., Haberl, F., Novara, G., Salvaterra, R., Turolla, R., Watson, M., Wilms, J., \& Wolter, A.} 2017,
\textit{Science}, 355, 817

\bibitem[Kong \& Di Stefano (2003)]{KongDiStefano2003}
{Kong, A.K.H., \& Di Stefano, R.} 2003,
\textit{ApJ} (Letters), 590, L13

\bibitem[Shakura \& Sunyaev (1973)]{ShakuraSunyaev1973}
{Shakura, N.I., \& Sunyaev, R.A.} 1973,
\textit{A\&A}, 24, 337

\bibitem[Tsygankov \etal\ (2016)]{Tsygankov_etal2016}
{Tsygankov, S.S., Mushtukov, A.A., Suleimanov, V.F., \& Poutanen, J.} 2016,
\textit{MNRAS}, 457, 1101

\end{thebibliography}
\end{document}